\documentclass[11pt]{article}
\pdfoutput=1
\usepackage{graphicx,float,multirow,hhline,fancyvrb,graphics,indentfirst,varioref,wrapfig,lettrine,subfigure}
\usepackage{soul}
\usepackage{amsmath}
\usepackage{amssymb}
\usepackage[latin1]{inputenc}
\usepackage{enumerate}
\usepackage{epstopdf}
\usepackage{epsfig}
\usepackage{graphicx}
\usepackage{color}
\usepackage{subfigure}
\usepackage{IEEEtrantools}
\usepackage{lineno}
\usepackage{wrapfig}
\usepackage{cite}

\DeclareGraphicsExtensions{.pdf,.eps,.jpg,.png}

\usepackage{setspace}

\usepackage{authblk}

\makeglossary

\begin{document}


\title{Modeling the Aerodynamic Lift Produced by Oscillating Airfoils at Low Reynolds Number}

\author[1]{Muhammad Saif Ullah Khalid\thanks{Corresponding Author, m.saifullahkhalid@ceme.nust.edu.pk}}
\author[1]{Imran Akhtar\thanks{Adjunct Research Member, Interdisciplinary Center for
Applied Mathematics, MC0531, Virginia Tech, Blacksburg VA 24061,
USA.}} \affil[1]{Department of Mechanical Engineering \\ NUST
College of Electrical \& Mechanical Engineering \\ National
University of Sciences \& Technology \\ Islamabad, 44000, Pakistan}




\maketitle


\section*{Abstract}
{For present study, setting Strouhal Number (St) as control
parameter, numerical simulations for flow past oscillating NACA-0012
airfoil at $10^3$ Reynolds Numbers (Re) are performed. Temporal
profiles of unsteady forces; lift and thrust, and their spectral
analysis clearly indicate the solution to be a period-1 attractor
for low Strouhal numbers. This study reveals that aerodynamic forces
produced by plunging airfoil are independent of initial kinematic
conditions of airfoil that proves the existence of limit cycle.
Frequencies present in the oscillating lift force are composed of
fundamental ($f_s$), even and odd harmonics ($3{f_s}$) at higher
Strouhal numbers. Using numerical simulations, shedding frequencies
($f_s$) were observed to be nearly equal to the excitation
frequencies in all the cases. Unsteady lift force generated due to
the plunging airfoil is modeled by modified van der Pol oscillator.
Using method of multiple scales and spectral analysis of
steady-state CFD solutions, frequencies and damping terms in the van
der Pol oscillator model are estimated. We prove the applicability
of this model to all planar motions of airfoil; plunging, pitching
and flapping. An important aspect of currently-proposed model is
capturing the time-averaged value of aerodynamic lift coefficient.}

\section*{Keywords} Reduced-Order Modeling, Nonlinear Dynamics, Low Reynolds
Number Aerodynamics, Oscillating Airfoils, Limit Cycle

\section{Introduction}
To propose efficient and better designs for small swimming and
flying unmanned vehicles, understanding of the unsteady mechanisms
to generate lift and thrust forces at very low Reynolds numbers (Re)
is of key importance. Being a highly nonlinear system, fluid flowing
over these vehicles carries great complexities. Since Knoller
\cite{Knoller1909} and Betz \cite{Betz1912} revealed the mechanisms
for production of thrust due to vortex pattern behind oscillating
and pitching airfoils, respectively, this field has attracted the
attention of many researchers around the globe. Recently, due to the
advent of biomimicking flying (micro air vehicles) and swimming
robots (underwater vehicles), many research efforts in this
direction are being presented. Ho et al. \cite{Ho2003},
Triantafyllou et al. \cite{Triantafyllou2004}, Wang \cite{Wang2005},
Lehmann \cite{Lehmann2008} and Shyy et al. \cite{Shyy2010} have
provided detailed reviews regarding progress in the field of
unsteady aerodynamics for flapping flight of insects, birds and
robots. Study of mechanism for generation of unsteady forces is
still to be understood completely due to a wide spectrum of
parameters that are involved. Both the experimental and currently
available numerical techniques require costly resources involving a
large amount in terms of time and money. Considering this fact,
researchers have also focussed towards development of the reduced
order models. These models are based upon the reduced number of
states for a dynamical system. These models can be built by either
phenomenological or direct model-reduction approaches. For
phenomenological modeling technique, the behavior of a parameter
from the response of a physical system is modeled by a set of
ordinary differential equations. Self-excited oscillators are great
examples of this type covering the range from electric circuits to
the aero-elastic phenomena. On the contrary, a direct approach
requires computation of coherent structures in the flow to develop a
reduced-order model \cite{Sirovich1987}. Proper-orthogonal
decomposition (POD) based reduced-order models are classical
examples in this category. It forms the reduced-order basis
functions from the flow snapshots capturing optimal energy of a
dynamical system. The governing equations, such as the Navier-Stokes
equations, are projected onto these reduced basis to form a
reduced-order model thereby reducing the system from millions of
degrees of freedom to order of ten. These models have been
successfully employed for flow control, design, optimization, and
uncertainty quantification \cite{Akhtar2008}. In the phenomena-based
approach for the development of reduced order models, work by Skop
et al. \cite{Skop1973} is of vital importance. They introduced the
Skop-Griffin parameter in order to propose a modified van der Pol
oscillator model to predict the lift of a cylinder coupled with a
linear model for mathematical representation of structure's motion.
This model is a center of focus in the study of vortex-induced
vibrations. In Ref. \cite{Skop1975}, they extended their model from
rigid to elastic cylinders. Identifying the cubic nonlinearity in
lift force and quadratic relation between lift and drag forces for a
circular cylinder, Nayfeh et al. \cite{Nayfeh2003, Nayfeh2005} used
method of multiple scales \cite{Nayfeh1993, Nayfeh1995, Nayfeh2000}
and proposed the first-order accurate reduced order self-excited
oscillator models for steady and transient parts of the aerodynamic
forces. Following the higher order spectral analysis technique
\cite{Fung1998, Hajj2000} for the identification of nonlinear
parameters, Qin \cite{Qin2004} developed forced van der Pol
oscillator models for the rotational, inline and transversal
oscillations of circular cylinders considering primary resonance
with soft and hard excitation cases at $\mbox{Re}=10^4$. Janajreh et
al. \cite{Janajreh2008} extended this study for rotational
oscillations of circular cylinders at $\mbox{Re}=10^2$. Marzouk et
al. \cite{Marzouk2007} presented a second order accurate model for
the steady-state lift and drag forces for stationary circular
cylinders at different Re values. Keeping eccentricity as the
control parameter and using method of harmonic balance, Akhtar et
al. \cite{Akhtar2009A} performed the numerical simulations for flow
over the elliptical structures and approximated the transient and
steady state behavior of the lift and drag forces using a combined
van der Pol-Duffing oscillator model. These afore-mentioned studies
help enhance our understanding regarding the nonlinear mechanism for
production of the unsteady aerodynamic forces by the bluff bodies.
To extend this technique for the aerodynamic bodies, Ellenrieder
\cite{Ellenrieder2006} adopted the methodology proposed by Skop et
al. \cite{Skop1997} to present following oscillator model for
time-dependent lift produced by the airfoils undergoing forced
plunging motion;
\begin{align}
\ddot{C_L}-{{\omega}_s}G({{C_{L_o}}^2}-4{Q^2}){\dot{C_L}}+{{\omega_s}^2}{C_L}={\omega_s}F{\dot{h}}
\label{eqn:heavemodel}
\end{align}
\noindent where $\omega_s$ is the fundamental vortex shedding
frequency, $C_{L_o}$ is the maximum amplitude of the time-dependent
lift force coefficient, $h$ is an instantaneous position of airfoil
while plunging and, $G$ and $F$ are the constants determined from
the empirical formulation proposed by Skop et al. \cite{Skop1997}.
In the current study, we first show the existence of limit-cycle
behavior in the response of an airfoil performing forced oscillatory
motion. Here, the response of this nonlinear system is determined in
terms of lift and thrust forces. We consider different sets of
initial kinematic conditions of the oscillating airfoil. Identifying
the nonlinearities in the response, we present a modified forced van
der Pol oscillator model for lift-force coefficient. The analytical
solution of this nonlinear mathematical model is derived using
method of multiple scales; a powerful perturbation technique. The
strength of this technique to solve the perturbation problems lies
in the fact that both the small and large values of system's states
can be handled by the involvement of slow and fast time-scales. This
reduced-order model not only captures the temporal details of the
the lift force but also it predicts its spectral composition
accurately. The purpose of this study is to provide a computational
tool for estimating the response of this physical system quickly. To
determine the parameters of the presented reduced-order model, we
employ numerical data obtained from CFD (computational fluid
dynamics) simulations using ANSYS Fluent \cite{Ansys}. We perform
these CFD simulations for a range of St values at $Re=10^3$. These
simulations are thoroughly validated and used here as the first step
towards development of the reduced-order models. The manuscript is
organized as follows. Section~\ref{sec:nummethd} provides necessary
details on our CFD simulation methodology for flow past the
oscillating airfoils along with the validation studies. Using
different sets of the kinematic initial conditions and states of
this nonlinear system in section~\ref{sec:lmtcycle}, we show the
limit-cycle existence in the response of the oscillation airfoils.
Mathematical formulation and derivation of a reduced-order model
based on a modified forced van der Pol oscillator equation is
presented in section~\ref{sec:rom}. Applicability of this model for
lift produced by plunging, pitching and flapping airfoils is shown
in section~\ref{sec:resultrom}. This model is capable of accurately
predicting the nonlinear behavior of unsteady $C_L$ of oscillating
airfoils. To model its time-averaged value $\bar{C_L}$, we introduce
a different type of quadratic nonlinearity to the self-excited van
der Pol oscillator model and describe its details in
section~\ref{sec:imprmod}. Section~\ref{sec:prdctv} explains the
overall behavior of the linear and nonlinear damping terms with
respect to increasing control parameter; the Strouhal number
(defined in section~\ref{sec:nummethd}). We also show the results of
predictive settings for the present model to compute unsteady $C_L$
and its spectra for the intermediate values of Strouhal number where
the model parameters were calculated through interpolation of the
data-set from CFD solutions. In section\ref{sec:cnclsion}, we
conclude and summarize our present work.
\section{Numerical Methodology}
\label{sec:nummethd} 
\begin{figure}[!ht]
\centering
\includegraphics[scale=0.35]{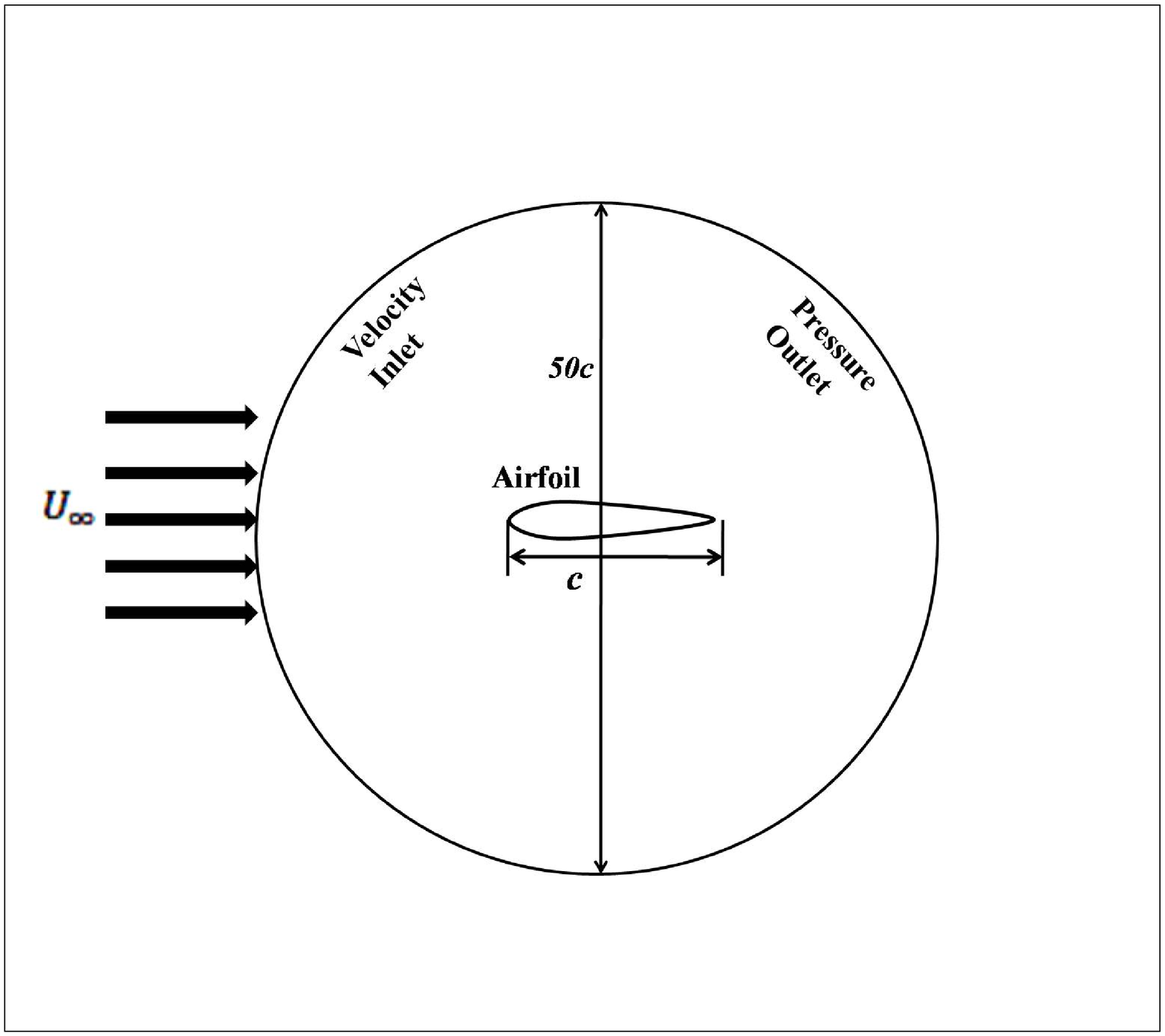}
\caption{Schematic of Geometry and Fluid-Domain} \label{fig:geom}
\end{figure}

\begin{figure}[!ht]
\centering
\includegraphics[scale=0.5]{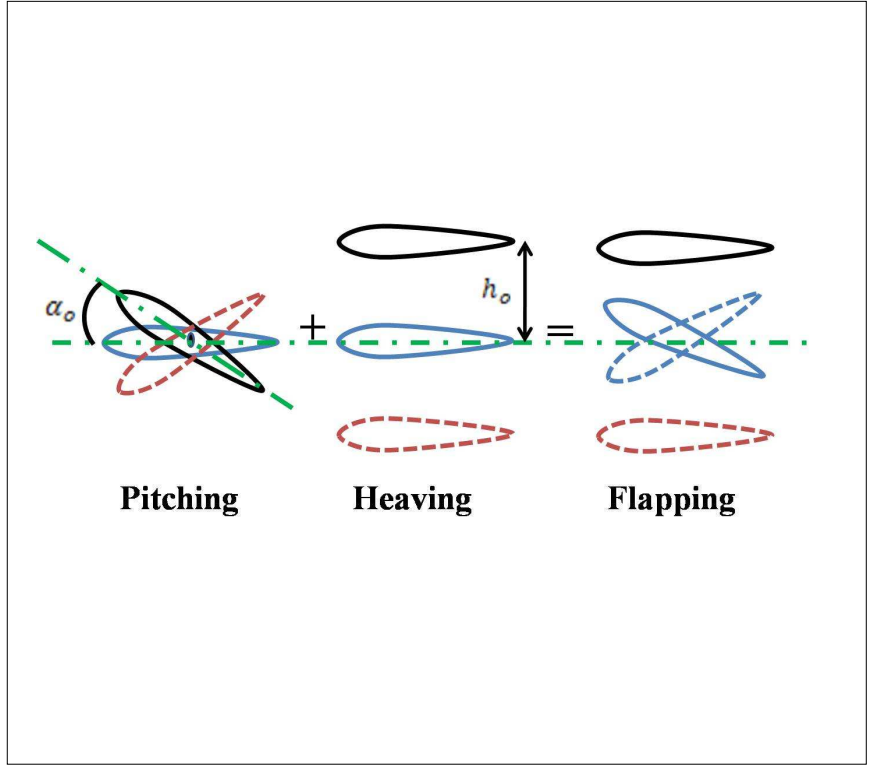}
\caption{Airfoil Kinematics (Solid Lines: Upstroke, Dotted Lines:
Downstroke) (For clarity, only mean and the amplitude positions are
shown here)} \label{fig:foilkinem}
\end{figure}

For the present study, we simulate flow past an
oscillating NACA-0012 airfoils by solving two-dimensional
incompressible Navier-Stokes equations using ANSYS-Fluent; a
finite-volume based commercial software. The Navier-Stokes equations
in their integral form can be written as;
\begin{align}
{\int\limits_V}{\frac{\partial{\rho}{\phi}}{\partial{t}}}{dV}+{\oint}{\rho}{\phi}{\vec{v}}.{d{\vec{A}}}-{\oint{\Gamma_{\phi}}}{\nabla}{\phi}.{d{\vec{A}}}-{\int
\limits_V{S_{\phi}}}{dV}=0
\end{align}
\noindent where $\rho$ is the density of fluid, $\vec{v}$ represents
the velocity vector, $\vec{A}$ shows the surface-area vector,
$\Gamma_{\phi}$ is the diffusion term, $\nabla{\phi}$ denotes the
gradient term and $S_{\phi}$ shows the source term. Since the
present case employs moving mesh technique \cite{Ansys}, the source
term is zero. The temporal term is approximated using the
first-order implicit method. Second-order upwind scheme is employed
for numerically approximating the convective term. Assuming
incompressible flow presently, velocity and pressure terms are
coupled through the pressure implicit with splitting of operator
(PISO) algorithm.
To avoid effect of disturbances on boundaries,
radius of the O-type domain is kept at $25c$, where c is the
chord-length of airfoil, as shown in Fig.~\ref{fig:geom}. Flow
domain is meshed using unstructured triangular cells. We use high
grid resolution near the airfoil surface to resolve the boundary
layer and to capture the wake characteristics downstream of the
airfoil. Airfoil surface is resolved using $400$ nodes. Dynamic
meshing techniques; spring analogy and remeshing, are employed in
the vicinity of moving airfoil for grid transition that allows the
adjustment of the grid in accordance with the instantaneous position
of airfoil.
Numerical solutions of flow-fields are highly dependent on the
suitability and accuracy of boundary conditions. In Fluent, motion
of an object may be defined by a user-defined function (\emph{UDF})
which is a computer code written in the C-language environment
coupled with the Fluent-Macros. Forced motion of an airfoil can be
of three types; plunging, pitching and flapping (combination of
plunging and pitching). The schematic of these motion are shown in
Fig.~\ref{fig:foilkinem}. Pithing and plunging motion can be modeled
as;
\begin{equation}
\begin{aligned}
Plunging: h(t) &= {h_{\circ}}\cos(2{\pi}{f}{t}+{\phi_h})\\
Pitching: \alpha(t) &= {\alpha_{\circ}}\sin(2{\pi}{f}{t})
\end{aligned}
\end{equation}
\noindent where $\alpha_{\circ}$ is the maximum pitching amplitude
and $\phi_h$ denotes the phase-angle.
Dirichlet conditions are employed on the inlet boundary and the
pressure outlet condition is used for the outflow boundary. At this
boundary, static pressure is specified. For incompressible flows,
pressure on the boundary is determined by taking the average of
specified values on the cell faces and computed values of the static
pressure on the corresponding cell-centers. All other flow variables
are computed by extrapolation of the computed values from the inner
domain. The Reynolds number, defined as ${U_\infty}{c}/{\nu}$, is
calculated using appropriate values of the kinematic viscosity
$\nu$, keeping all other parameters equal to unity. Strouhal number
(St) is defined as $\mbox{St}={2f{h_{\circ}}}/{U_\infty}$ where
$h_{\circ}$ is the maximum amplitude, {$f$} is the excitation
frequency for oscillating airfoil in Hertz, and $U_\infty$ is the
free-stream fluid velocity. It is considered as the primary
governing parameter for investigating unsteady behavior of
oscillating airfoils. $h_\circ$ depicts the wake-width behind a
plunging airfoil. For pitching motion, total vertical excursion
traversed by the trailing-edge of airfoil approximates the
wake-width. St is varied by changing plunging and pitching
amplitudes while incoming reference flow velocity and oscillation
frequency are kept fixed. In the present study, numerical
simulations are initialized by the uniform inlet velocity boundary
conditions. To perform these simulations, we employ
$1.18440\times10^5$ number of cells in the whole domain and $2000$
time-steps per oscillation cycle of the airfoil. Details for the
grid-convergence, time-step refinement, and validation studies are
available in Ref. \cite{KhalidJoA2014}. Lift and thrust forces are
computed by integrating pressure and shear stresses over the surface
of airfoil. Their corresponding coefficients, $C_L$ and $C_T$
respectively are computed as;
\begin{equation}
\begin{aligned}
C_L &= \frac{L}{{\frac{1}{2}}{\rho}{{U_\infty}^2}{c}} \\
C_T &= \frac{T}{{\frac{1}{2}}{\rho}{{U_\infty}^2}{c}}
\end{aligned}
\end{equation}
Using time-period of an oscillation cycle ($\tau=1/f$),
corresponding time-averaged coefficients are calculated through
following relation;
\begin{align}
\bar{C}={\frac{1}{\tau}}{\int_t^{t+\tau} {C(t)} dt}
\end{align}
\section{Existence of Limit-Cycle}
\label{sec:lmtcycle} 
\begin{figure}[!ht]
\centering
\includegraphics[scale=0.5]{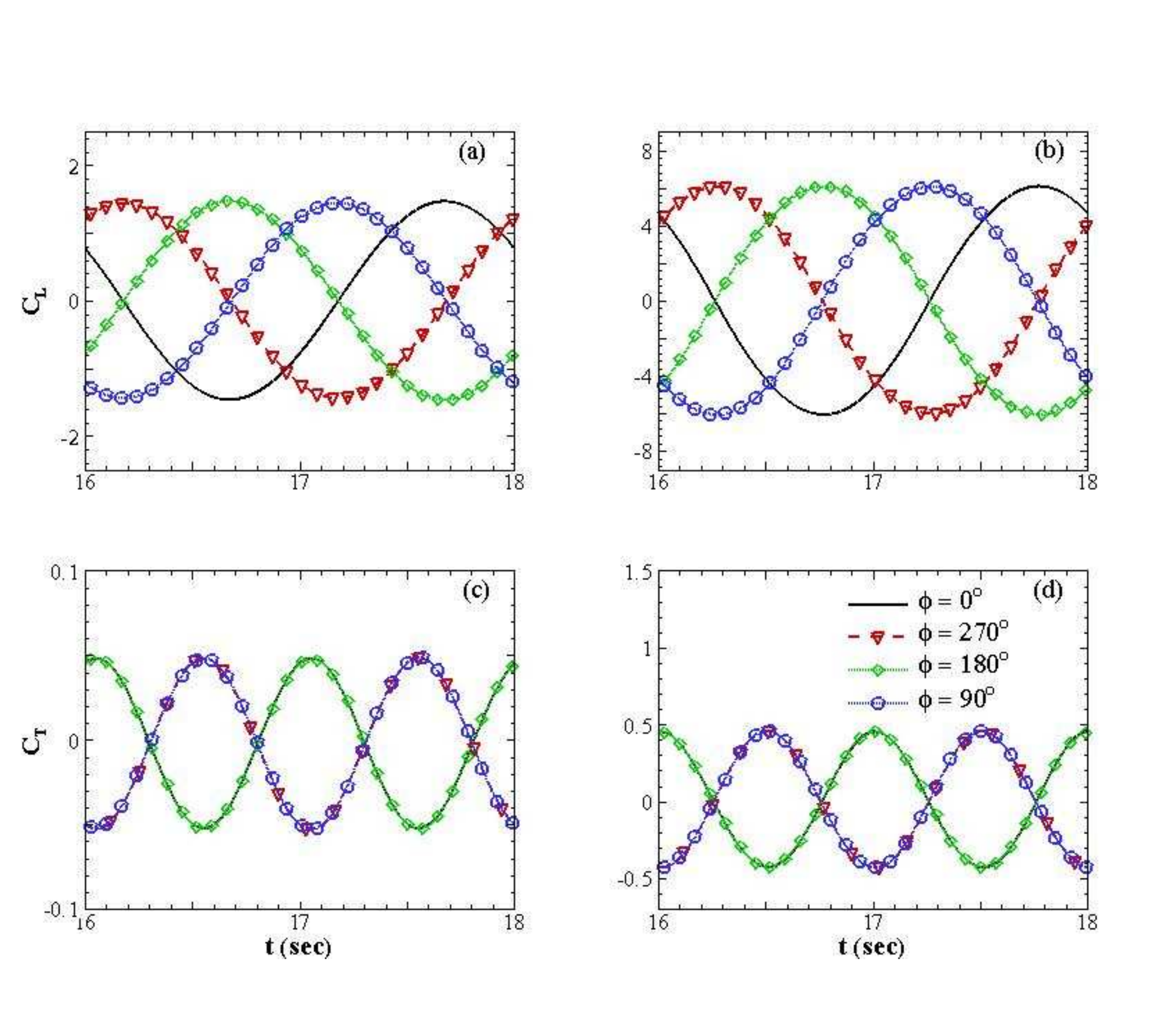}
\caption{Comparison of $C_L$ and $C_T$ for different initial
positions and strokes of plunging airfoil for $\mbox{St}=0.10$ in
(a) and (c); and for $\mbox{St}=0.30$ in (b) and (d)}
\label{fig:CLCTCMPRSN}
\end{figure}

\begin{figure}[!ht]
\centering
\includegraphics[scale=0.5]{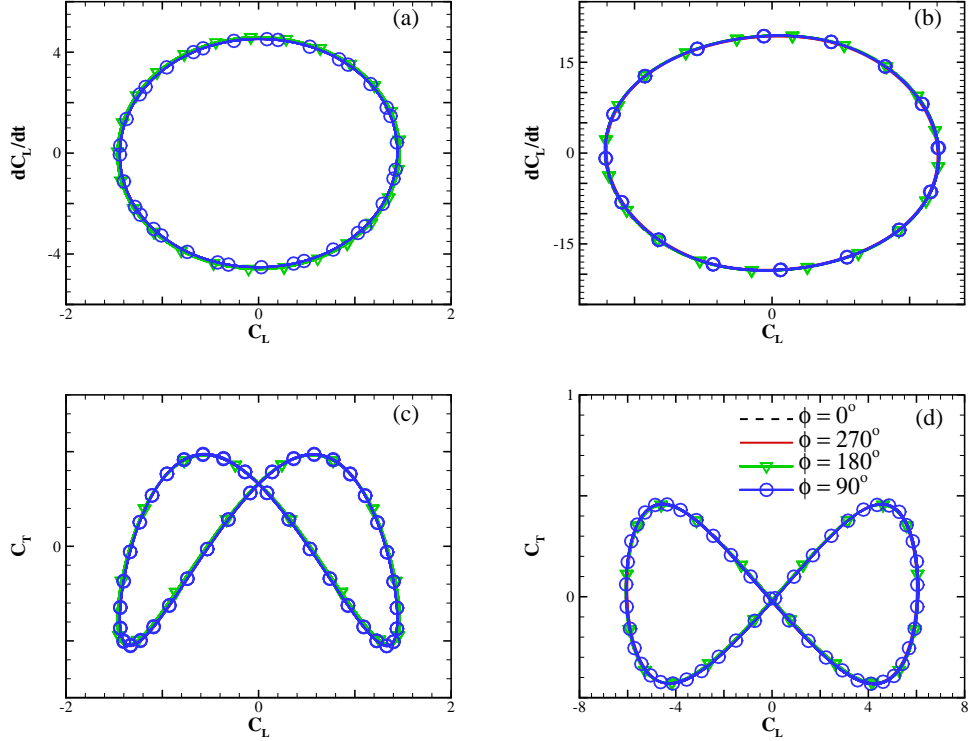}
\caption{Phase maps using $C_L$ and $C_T$ their time-derivatives for
different initial positions and strokes of plunging airfoil at
$\mbox{St}=0.10$ in (a) and (c); and for $\mbox{St}=0.30$ in (b) and
(d)} \label{fig:CLCTPhase}
\end{figure}

Limit cycles can be used to model various
nonlinear oscillatory systems from the real world. Limit-cycle
exists if energy input to the dynamical system is balanced by an
energy dissipation mechanism and the system's response limits itself
onto either a periodic, a quasi-periodic, or a chaotic attractor
\cite{Mn1998}. Hence this phenomenon represents a closed trajectory
in the phase space. To characterize the behavior of a nonlinear
system as a limit cycle, it needs to be tested for various initial
conditions. Based upon a reduced-order model for $C_L$, von
Ellenrieder \cite{Ellenrieder2006} and von Ellenrieder et al.
\cite{Ellenrieder2008} presented the aerodynamic response of
plunging airfoil as a limit cycle. To obtain numerical solution for
governing ordinary differential equation, all the initial conditions
were described in the form of $C_L$ and its time-derivatives.
Use of
the actual initial kinematic conditions seems more appropriate to
investigate the limit-cycle behavior of any dynamical system. For
current research-work, we consider a broader spectrum of the initial
conditions through inclusion of initial position where the airfoil
starts its motion from. There can be infinite starting positions
between positive and negative amplitudes of an oscillating airfoil.
In this study, initial conditions refer to the initial kinematic
states of plunging motion for the numerical simulations. These
include following scenarios;
\begin{enumerate}
\item{Does plunging airfoil undergo upstroke or downstroke first?}
\item{Which position does it start its oscillation from?}
\end{enumerate}
We choose three positions to start the airfoil's oscillatory motion.
These are the top-most, mean (mid) and the bottom-most positions.
Airfoil can perform upstroke initially starting from the bottom-most
or mean position and it may undergo downstroke starting from mean or
the top-most position. ${\phi_h}={0^\circ}$ gives initial downstroke
starting from the maximum positive amplitude of oscillating airfoil.
Using these conditions of initial stroke and position, four cases
are required to be studied for one set of values of oscillation
frequency and amplitude. To analyze this phenomena here, we consider
four initial conditions described earlier in terms of the starting
position and stroke of a plunging airfoil. We compute the response
of this dynamical system in terms of the aerodynamic lift and thrust
coefficients denoted as $C_L$ and $C_T$, respectively. Using the
temporal histories and phase maps of these dynamical states, we
observe similarity in their nonlinear character for different
initial conditions.
In Fig.~\ref{fig:CLCTCMPRSN}, $C_L$ and $C_T$
for four initial conditions of plunging airfoil are presented. We
observe the same phase difference in their temporal histories as
given in the forced motion of plunging airfoil. Another notable
feature is the same amplitude of $C_L$ for all the initial
conditions. Phase of $C_L$ is exactly equal to that of heave motion.
But for $C_T$, phase depends upon magnitude of $\phi$ only and not
its sign (where we can say $270^{\circ}=-90^{\circ}$).
Here, we support the existence of limit-cycle using aerodynamics
forces and their time-derivatives as the states of flow past a
plunging airfoil through phase maps shown in
Fig.~\ref{fig:CLCTPhase} for $\mbox{St}=0.10$ and $0.30$. Periodic
steady-state solutions for the aerodynamic force coefficients are
presented for this purpose. Like bluff-body aerodynamics
\cite{Nayfeh2003, Marzouk2007, Akhtar2009A}, the frequency of
unsteady $C_T$ is twice of that for the unsteady $C_L$. All the
initial conditions lead us to the same period-{\it n} attractor thus
proving independence of this system from initial kinematic states of
oscillating airfoil for the range of parameters considered here.
\section{Reduced Order Modeling}
\label{sec:rom} 
\begin{figure}[!ht]
\centering
\includegraphics[scale=0.5]{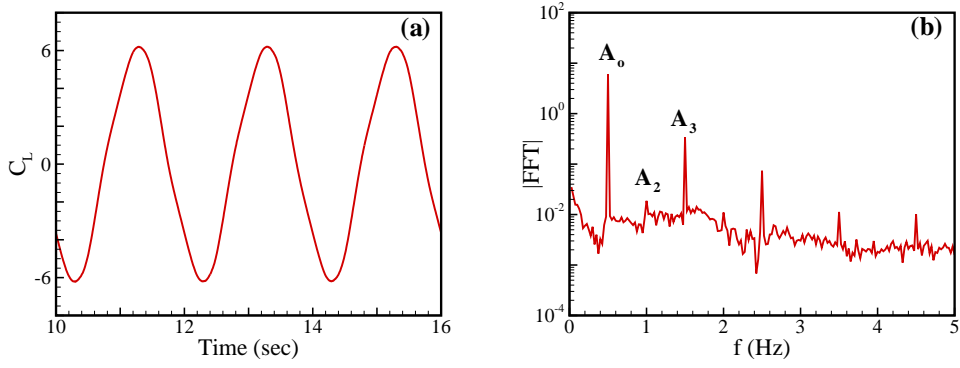}
\caption{(a) Unsteady Lift and (b) Amplitude-Spectra for
$\mbox{St}=0.30$} \label{fig:Idntf}
\end{figure}

Young et al. \cite{Young2007} explained the
existence of vortex-lock in phenomena for plunging airfoil for a
range of flow and kinematic parameters. In such cases, natural
vortex-shedding frequency comes out to be equal to forcing frequency
of oscillating airfoil. Presence of higher-order harmonics in
spectra of unsteady aerodynamic force coefficients for flow over
oscillating airfoils exhibits this phenomena as nonlinear. This is a
well-known fact in vibrations of bluff-bodies like cylinders and
cables as well \cite{Singh2008, Smith2012}.
In current study, spectra
of the unsteady lift force are analyzed to identify the prominent
frequency components and their corresponding harmonics present in
the signal. For given range of St, (from 0.05 to 0.5), fundamental
vortex shedding $f_s$ frequency appears to be equal to the forced
plunging frequency ($\Omega/{2\pi}$), equal to 0.5Hz of the airfoil.
It indicates that these all are the cases related to the primary
resonance. First even and odd harmonics show prominent peaks. As we
increase St, even harmonic starts getting larger amplitude.
Figure~\ref{fig:Idntf} shows three shedding cycles represented by
unsteady $C_L$ and its relevant amplitude-spectra. The amplitudes of
energetic fundamental, first even and odd harmonics are also
indicated being employed here for model-reduction.
An asymmetric forced van der Pol oscillator model is used here to
analytically represent the lift force produced by plunging NACA-0012
airfoil. Due to the presence of even harmonics in $C_L$-spectra, we
introduce a quadratic nonlinear term containing a multiple of
${C_L}{\dot{C_L}}$. Choice of this term to model the quadratic
nonlinearity is made due to a phase-difference of nearly $\pi/2$ (or
its odd integral multiples) between the fundamental and first even
harmonics. Assuming this system as weekly damped and having a soft
excitation;
\begin{align}
\ddot{C_L}+{{\omega^2}_s}{C_L} &=
{\epsilon}[{\mu}{\dot{C_L}}-{\alpha}{C_L}{\dot{C_L}}-{\gamma}{{{C_L}^2}{\dot{C_L}}+{F_\circ}
{\cos({\Omega}{t}+{\Gamma})}}] \label{eqn:vdp1}
\end{align}
\noindent $\omega_s$ is the vortex-shedding frequency, $\epsilon$ is
a book-keeping parameter while $\mu$, $\alpha$ and $\gamma$
represent the linear, quadratic and cubic damping coefficients,
respectively. All these parameters are positive real numbers.
Although no vortex-shedding from NACA-0012 is observed at zero
angle-of-attack at this $\mbox{Re}$, we consider this case related
to primary resonance ($\Omega \approx {\omega_s}$) in a mathematical
sense;
\begin{align}
\Omega = {\omega}+{\epsilon}{\sigma} \label{eqn:vdp2}
\end{align}
\noindent Using Eq.~\ref{eqn:vdp2} and ${F_\circ}$ as the amplitude,
excitation may be expressed as;
\begin{align}
E(t) = {{F_\circ}} \cos[({\omega_s}+\epsilon \sigma)t + \Gamma]
\label{eqn:vdp3}
\end{align}
\noindent Assuming time scales are ${T_{\circ}}=t$,
${T_1}={\epsilon}t$ and ${T_2}={{\epsilon}^2}t$, we have;
\begin{align}
E(t) = {F_\circ} \cos[({\omega_s}{T_{\circ}}+{\sigma}{T_1}+\Gamma]
\label{eqn:vdp4}
\end{align}
\noindent Using Chain rule for differentiation, we can write;
\begin{align}
\frac{d}{dt} &= {D_{\circ}}+{{\epsilon}{D_1}}+{{\epsilon
^2}{D_2}}+... \label{eqn:vdp5}
\end{align}
\begin{align}
\frac{d^2}{d{t^2}} = {{D^2}_{\circ}}+
{2{\epsilon}{D_{\circ}}{D_1}}+{{\epsilon}^2}({{D^2}_1}+{2{D_{\circ}}{D_1}})+...
\label{eqn:vdp6}
\end{align}
\noindent Let second-order approximate solution for
Eq.~\ref{eqn:vdp1} be;
\begin{align}
{C_L}(t) =
{{C_L}_{\circ}}{(T_{\circ},T_1,T_2)}+{\epsilon}{{C_L}_1}{(T_{\circ},T_1,T_2)}+{{\epsilon}^2}{{C_L}_2}{(T_{\circ},T_1,T_2)}
\label{eqn:vdp7}
\end{align}
\noindent We expand all the terms in Eq.~\ref{eqn:vdp1} as follows;
\begin{IEEEeqnarray}{rCl}
\ddot{C_L} &=& ({{D^2}_{\circ}}+{2{\epsilon}{D_{\circ}}{D_1}}+{{\epsilon ^2}{{D^2}_1}}+{2{\epsilon ^2}{D_{\circ}}{D_2}})({{C_L}_{\circ}}+{{\epsilon}{{C_L}_1}}+{{\epsilon ^2}{{C_L}_2}}) \nonumber\\
&& =\:
{{D^2}_{\circ}}{{C_L}_{\circ}}+{{\epsilon}{{D^2}_{\circ}}{{C_L}_1}}+{{\epsilon
^2}{{D^2}_{\circ}}{{C_L}_2}}+{2{\epsilon}{D_{\circ}}{D_1}{{C_L}_{\circ}}}+{2{\epsilon
^2}{D_{\circ}}{D_1}{{C_L}_1}}\nonumber\\
&&+\: {{\epsilon ^2}{{D^2}_1}{{C_L}_{\circ}}}+{2{\epsilon
^2}{D_{\circ}}{D_2}{{C_L}_{\circ}}} \label{eqn:vdp8}
\end{IEEEeqnarray}
\begin{IEEEeqnarray}{rCl}
{{\omega^2}_s}{C_L} &=& {{\omega ^2}_s}({{C_L}_{\circ}}+{{\epsilon}{{C_L}_1}}+{{\epsilon ^2}{{C_L}_2}}) \nonumber\\
&& =\: {{\omega ^2}_s}{{C_L}_{\circ}}+{\epsilon}{{\omega
^2}_s}{{C_L}_1}+{\epsilon ^2}{{\omega ^2}_s}{{C_L}_2}
\label{eqn:vdp9}
\end{IEEEeqnarray}
\begin{IEEEeqnarray}{rCl}
{\epsilon}{\mu}{\dot{C_L}} &=&  {\epsilon}{\mu}({D_{\circ}}+{{\epsilon}{D_1}}+{{\epsilon ^2}{D_2}}+...)({{C_L}_{\circ}}+{{\epsilon}{{C_L}_1}}+{{\epsilon}{{C_L}_2}})\nonumber\\
&& =\: {{\epsilon}{\mu}{D_{\circ}}{{C_L}_{\circ}}}+{{\epsilon
^2}{\mu}{D_1}{{C_L}_1}}+{{\epsilon ^2}{\mu}{D_1}{{C_L}_{\circ}}}
\label{eqn:vdp10}
\end{IEEEeqnarray}
\begin{IEEEeqnarray}{rCl}
-{\epsilon}{\alpha}{C_L}{\dot{C_L}} &=&  -{\epsilon}{\alpha}({{C_L}_{\circ}}+{{\epsilon}{{C_L}_1}}+{{\epsilon ^2}{{C_L}_2}})({D_{\circ}}+{{\epsilon}{D_1}}+{{\epsilon ^2}{D_1}})({{C_L}_{\circ}}+{{\epsilon}{{C_L}_1}}+{{\epsilon ^2}{{C_L}_2}})\nonumber\\
&& =\:
-{{\epsilon}{\alpha}{{C_L}_{\circ}}({D_{\circ}}{{C_L}_{\circ}}}) -
{{\epsilon ^2}{\alpha}{{C_L}_{\circ}}({D_{\circ}}{{C_L}_1})} \nonumber\\
&&-\: {{\epsilon ^2}{\alpha}{{C_L}_{\circ}}({D_1}{{C_L}_{\circ}})} -
{{\epsilon
^2}{\alpha}{{C_L}_1}({D_{\circ}}{{C_L}_{\circ}})}\label{eqn:vdp11}
\end{IEEEeqnarray}
\begin{IEEEeqnarray}{rCl}
-{\epsilon}{\gamma}{\dot{C_L}}{{C_L}^2} &=& -{\epsilon}{\gamma}{{[{{C_L}_{\circ}}+{{\epsilon}{{C_L}_1}}+{{\epsilon ^2}{{C_L}_2}}]}^2}{({D_{\circ}}+{{\epsilon}{D_1}}+{{\epsilon ^2}{D_1}})}({{{C_L}_{\circ}}+{{\epsilon}{{C_L}_1}}+{{\epsilon ^2}{{C_L}_2}}}) \nonumber\\
&& =\:
-{\epsilon}{\gamma}{{{C_L}^2}_{\circ}}({D_{\circ}}{{C_L}_{\circ}}) -
{{\epsilon
^2}{\gamma}{{{C_L}^2}_{\circ}}({D_{\circ}}{{C_L}_{\circ}})} \nonumber\\
&&-\: {{\epsilon
^2}{\gamma}{{{C_L}^2}_{\circ}}({D_1}{{C_L}_{\circ}})} - {2{\epsilon
^2}{\gamma}{{C_L}_{\circ}}{{C_L}_1}({D_{\circ}}{{C_L}_{\circ}})}
\label{eqn:vdp12}
\end{IEEEeqnarray}
\noindent
${\epsilon}{F_\circ}{\cos({{{\omega}_s}{T_{\circ}}}+{{\sigma}{T_1}}+{\Gamma})}$
remains in its present form. Equating like powers of $\epsilon$ on
both sides of Eq.~\ref{eqn:vdp1}, we get;
\begin{align}
\mathcal{O}(\epsilon^{\circ}): \quad {{D^2}_{\circ}}{{C_L}_{\circ}}
+ {{\omega}_s}{{C_L}_{\circ}} = 0 \label{eqn:vdp13}
\end{align}
\begin{IEEEeqnarray}{rCl}
\mathcal{O}(\epsilon^1): \quad {{D^2}_{\circ}}{{C_L}_1} + {{\omega}_s}{{C_L}_1} &=& -{2{D_{\circ}}{D_1}{{C_L}_{\circ}}} + {{\mu}{D_{\circ}}{{C_L}_{\circ}}} - {{\alpha}{{C_L}_{\circ}}({D_{\circ}}{{C_L}_{\circ}})} - {{\gamma}{{{C_L}^2}_{\circ}}({D_{\circ}}{{C_L}_{\circ}})} \nonumber\\
&& +\:
{{F_\circ}\cos({{\omega}_s}{T_{\circ}}+{\sigma}{T_1}+{\Gamma})}
\label{eqn:vdp14}
\end{IEEEeqnarray}
\begin{IEEEeqnarray}{rCl}
\mathcal{O}(\epsilon^2): \quad {{D^2}_{\circ}}{{C_L}_2} + {{\omega}_s}{{C_L}_2} &=& - {2{D_{\circ}}{D_1}{{C_L}_1}} - {{{D^2}_1}{{C_L}_{\circ}}} - {2{D_{\circ}}{D_2}{{C_L}_{\circ}}} + {{\mu}({D_{\circ}}{{C_L}_1})} - {{\alpha}{{C_L}_{\circ}}({D_{\circ}}{{C_L}_1})} \nonumber\\
&& -\: {{\alpha}{{C_L}_{\circ}}({D_1}{{C_L}_{\circ}})} -
{{\alpha}{{C_L}_1}({D_{\circ}}{{C_L}_{\circ}})} \nonumber\\
&& -\: {{\gamma}{{{C_L}^2}_{\circ}}({D_{\circ}}{{C_L}_1})} -
{{\gamma}{{{C_L}^2}_{\circ}}({D_1}{{C_L}_{\circ}})} -
{2{\gamma}{{C_L}_{\circ}}{{C_L}_1}({D_{\circ}}{{C_L}_{\circ}})}
{}\label{eqn:vdp15}
\end{IEEEeqnarray}
\noindent Solution of Eq.~\ref{eqn:vdp13} is;
\begin{align}
{C_L}_{\circ}(t) =
{A_{\circ}}(T_1,T_2){\cos[{{\omega}_s}{T_{\circ}}+{\theta_{\circ}}(T_1,T_2)]}
\label{eqn:vdp16}
\end{align}
\noindent To solve Eq.~\ref{eqn:vdp14}, all the terms on right hand
side need expansion.
\begin{align}
-2
{\frac{{\partial^2}{{C_L}_{\circ}}}{{{\partial}{T_{\circ}}}{\partial{T_1}}}}
=
2{{\omega}_s}{\frac{\partial{A_{\circ}}}{\partial{T_1}}}{\sin({{\omega}_s}{T_{\circ}}+{\theta_{\circ}})}+2{{\omega}_s}{A_{\circ}}{\frac{\partial{\theta_{\circ}}}{\partial{T_1}}}{\cos({{\omega}_s}{T_{\circ}}+{\theta_{\circ}})}
\label{eqn:vdp17}
\end{align}
\begin{align}
{\mu}{\frac{\partial{{C_L}_{\circ}}}{\partial{T_{\circ}}}}=-{\mu}{{\omega}_s}{A_{\circ}}\sin({{\omega}_s}{T_{\circ}}+{\theta_{\circ}})
\label{eqn:vdp18}
\end{align}
\begin{align}
-{\alpha}{{C_L}_{\circ}}{\frac{\partial{{C_L}_{\circ}}}{\partial{T_{\circ}}}}={\frac{1}{2}}{\alpha}{{\omega}_s}{{A_{\circ}}^2}\sin(2{{\omega}_s}{T_{\circ}}+{2{\theta_{\circ}}})
\label{eqn:vdp19}
\end{align}
\begin{align}
-{\gamma}{{{C_L}_{\circ}}^2}{\frac{\partial{{C_L}_{\circ}}}{\partial{T_{\circ}}}}={{\frac{1}{4}}{\gamma}{{\omega}_s}{{A_{\circ}}^3}\sin({{\omega}_s}{T_{\circ}}+{\theta_{\circ}})}
+
{{\frac{1}{4}}{\gamma}{{\omega}_s}{{A_{\circ}}^3}{}\sin(3{{\omega}_s}{T_{\circ}}+3{\theta_{\circ}})}
\label{eqn:vdp20}
\end{align}
\begin{align}
{F_\circ}
\cos({{\omega}_s}{T_{\circ}}+{\sigma}{T_1}+{\Gamma})={F_\circ}
\cos({{\omega}_s}{T_{\circ}}+{\theta_{\circ}}+{\sigma}{T_1}+{\Gamma}-{\theta_{\circ}})\\
= {F_\circ}
\cos({{\omega}_s}{T_{\circ}}+{\theta_{\circ}})\cos({\sigma}{T_1}+{\Gamma}-{\theta_{\circ}})
- {F_\circ}
\sin({{\omega}_s}{T_{\circ}}+{\theta_{\circ}})\sin({\sigma}{T_1}+{\Gamma}-{\theta_{\circ}})
\label{eqn:vdp21}
\end{align}
\noindent Modulation equation may be obtained by combining multiples
of $\sin({{\omega}_s}{T_{\circ}}+{\theta_{\circ}})$;
\begin{align}
{2{{\omega}_s}{\frac{\partial{A_{\circ}}}{\partial{T_1}}}}-{{\mu}{{\omega}_s}{A_{\circ}}}+{\frac{{\gamma}{{\omega}_s}{{A_{\circ}}^3}}{4}}-{{F_\circ}\sin({{\sigma}{T_1}}{\Gamma}{\theta_{\circ}})}=0
\label{eqn:vdp22}
\end{align}
\begin{align}
\frac{\partial{A_{\circ}}}{\partial{T_1}}={\frac{{\mu}{A_{\circ}}}{2}}-{\frac{{\gamma}{{A_{\circ}}^3}}{8}}-{{\frac{{F_\circ}}{{2}{{\omega}_s}}}{\sin({{\sigma}{T_1}}+{\Gamma}+{\theta_{\circ}})}}
\label{eqn:vdp23}
\end{align}
\noindent By combining multiples of
$\cos({{\omega}_s}{T_{\circ}}+{\theta_{\circ}})$, second modulation
equation may be constructed.
\begin{align}
2{{\omega}_s}{A_{\circ}}{\frac{\partial{\theta_{\circ}}}{\partial{T_1}}}={F_\circ}
\cos({{\sigma}{T_1}}+{\Gamma}-{\theta_{\circ}}) \label{eqn:vdp24}
\end{align}
\noindent To make the system autonomous, we assume;
\begin{equation*}
\begin{aligned}
\eta = {{\sigma}{T_1}}+{\Gamma}-{\theta_{\circ}} \\
\theta_{\circ} = {{\sigma}{T_1}}+{\Gamma}-{\eta} \\
\frac{\partial{\theta_{\circ}}}{\partial{T_1}} = {\sigma} -
{\frac{\partial{\eta}}{\partial{T_1}}}
\end{aligned}
\end{equation*}
\noindent So, Eq.~\ref{eqn:vdp24} may be written as;
\begin{align}
\frac{\partial{\eta}}{\partial{T_1}} = {\sigma} +
{\frac{{F_\circ}}{{2}{{\omega}_s}{A_{\circ}}}}{\cos(\eta)}
\label{eqn:vdp28}
\end{align}
\noindent Thus amplitude and phase are governed by;
\begin{align}
\dot{A_{\circ}} =
{\frac{{\mu}{A_{\circ}}}{2}}-{\frac{{\gamma}{{A_{\circ}}^3}}{8}}-{{\frac{{F_\circ}}{{2}{{\omega}_s}}}{\sin({{\sigma}{T_1}}+{\Gamma}+{\theta_{\circ}})}}
\label{eqn:vdp29}
\end{align}
\begin{align}
\dot{\eta}
={\sigma}+{\frac{{F_\circ}}{{2}{{\omega}_s}{A_{\circ}}}}{\cos(\eta)}
\label{eqn:vdp30}
\end{align}
\noindent To get the steady-state motion, time-derivative of both
amplitude and phase needs to be zero.
\begin{align}
{\frac{{\mu}{A_{\circ}}}{2}}-{\frac{{\gamma}{{A_{\circ}}^3}}{8}}={{\frac{{F_\circ}}{{2}{{\omega}_s}}}{\sin({{\sigma}{T_1}}+{\Gamma}+{\theta_{\circ}})}}
\label{eqn:vdp31}
\end{align}
\begin{align}
{\sigma}{A_{\circ}}={\frac{{F_\circ}}{{2}{{\omega}_s}}}{\cos(\eta)}
\label{eqn:vdp32}
\end{align}
\noindent Squaring and adding both Eq.~\ref{eqn:vdp31} and
\ref{eqn:vdp32}, we get;
\begin{align}
\frac{{F_\circ}^2}{{4}{{{\omega}_s}^2}} =
{(-{\frac{{\mu}{A_{\circ}}}{2}}+{\frac{{\alpha}{{A_{\circ}}^3}}{8}})^2}
+ {{{A_{\circ}}^2}{\sigma ^2}} \label{eqn:vdp33}
\end{align}
\noindent Identifying the modulating terms, we get following
governing expression for ${C_L}_1(t)$;
\begin{align}
\frac{\partial^2{{C_L}_1}}{\partial{{T_{\circ}}^2}} +
{{{\omega}_s}^2}{{C_L}_1} =
{{\alpha}{{\omega}_s}{\frac{{A_{\circ}}^2}{2}}{\sin({{2}{{\omega}_s}{T_{\circ}}}+{{2}{\theta_{\circ}}})}}
+
{{\gamma}{{\omega}_s}{\frac{{A_{\circ}}^3}{4}}{\sin({{3}{{\omega}_s}{T_{\circ}}}+{{3}{\theta_{\circ}}})}}
\label{eqn:vdp34}
\end{align}
\noindent To get particular solution, we can write;
\begin{align}
{C_L}_{1p} = {K}
\sin({{2}{{\omega}_s}{T_{\circ}}}+{{2}{\theta_{\circ}}}) + {M}
\sin({{3}{{\omega}_s}{T_{\circ}}}+{{3}{\theta_{\circ}}})
\label{eqn:vdp35}
\end{align}
\begin{align}
\frac{\partial{{C_L}_{1p}}}{\partial{T_{\circ}}} =
{2}{{\omega}_s}{K}
\cos({{2}{{\omega}_s}{T_{\circ}}}+{{2}{\theta_{\circ}}}) +
{3}{{\omega}_s}{M}
\cos({{3}{{\omega}_s}{T_{\circ}}}+{{3}{\theta_{\circ}}})
\label{eqn:vdp36}
\end{align}
\begin{align}
\frac{{\partial^2}{{C_L}_{1p}}}{\partial{{T_{\circ}}^2}} =
-{4}{{{\omega}_s}^2}{K}
\sin({{2}{{\omega}_s}{T_{\circ}}}+{{2}{\theta_{\circ}}}) -
{9}{{{\omega}_s}^2}{M}
\sin({{3}{{\omega}_s}{T_{\circ}}}+{{3}{\theta_{\circ}}})
\label{eqn:vdp37}
\end{align}
\noindent putting these values of ${C_L}_{1p}$,
$\frac{\partial{{C_L}_{1p}}}{\partial{T_{\circ}}}$ and
$\frac{{\partial^2}{{C_L}_{1p}}}{\partial{{T_{\circ}}^2}}$ into
Eq.~\ref{eqn:vdp34} and equating like terms, we get
$K=-\frac{{\alpha}{{A_{\circ}}^2}}{{6}{{\omega}_s}}$  and $M=-
\frac{{\gamma}{{A_{\circ}}^3}}{{32}{{\omega}_s}}$. Hence,
${C_L}_1(t)$ comes out to be;
\begin{align}
{C_L}_{1p}(t) = -
{\frac{{\alpha}{{A_{\circ}}^2}}{{6}{{\omega}_s}}}\sin({2}{{{\omega}_s}{t}}+{{2}{\theta_{\circ}}})
-
{\frac{{\gamma}{{A_{\circ}}^3}}{{32}{{\omega}_s}}}\sin({3}{{{\omega}_s}{t}}+{{3}{\theta_{\circ}}})
\label{eqn:vdp39}
\end{align}
\noindent Thus complete solution for ${C_L}(t)$ is;
\begin{align}
{C_L}(t) = {A_{\circ}}\cos({{\omega}_s}{t}+{\theta{\circ}}) -
{\epsilon}{\frac{{\alpha}{{A_{\circ}}^2}}{{6}{{\omega}_s}}}\cos({2}{{{\omega}_s}{t}}+{{2}{\theta_{\circ}}}+{\frac{\pi}{2}})
-
{\epsilon^2}{\frac{{\gamma}{{A_{\circ}}^3}}{{32}{{\omega}_s}}}\cos({3}{{{\omega}_s}{t}}+{{3}{\theta_{\circ}}}+{\frac{\pi}{2}})+...
\label{eqn:vdp40}
\end{align}
\noindent For autonomous dynamical system in complex notation, we
have;
%
%
\begin{IEEEeqnarray}{rCl}
{C_L}(t) &=& {{\frac{A_{\circ}}{2}}[{e^{i({{\Omega}{t}}+{\Gamma}-{\eta})}}+{e^{-i({{\Omega}{t}}+{\Gamma}-{\eta})}}]} \nonumber\\
&& -\:
{{\frac{{\alpha}{{A_{\circ}}}^2}{{12}{{\omega}_s}}}[{e^{i({{2}{\Omega}{t}}+{2}{\Gamma}-{2}{\eta}+{\frac{\pi}{2}})}}+{e^{-i({{2}{\Omega}{t}}+{2}{\Gamma}-{2}{\eta}+{\frac{\pi}{2}})}}]}
\nonumber\\
&& -\:
{{\frac{{\gamma}{{A_{\circ}}}^3}{{64}{{\omega}_s}}}[{e^{i({{3}{\Omega}{t}}+{3}{\Gamma}-{3}{\eta}+{\frac{\pi}{2}})}}+{e^{-i({{3}{\Omega}{t}}+{3}{\Gamma}-{3}{\eta}+{\frac{\pi}{2}})}}]}
\label{eqn:vdp42}
\end{IEEEeqnarray}
Now, we describe the procedure for identification of linear and
nonlinear parameters. Performing the spectral analysis of unsteady
lift profiles, values for fundamental vortex shedding frequency
$\omega$, its corresponding amplitude $A_{\circ}$ and amplitudes
$A_2$ and $A_3$ of first even and odd harmonics, respectively can be
measured. Comparing the assumed approximate solution of van der Pol
oscillator model in Eq.~\ref{eqn:vdp7} with that from multiple
scales method in Eq.~\ref{eqn:vdp40}, nonlinear damping parameters;
$\alpha$ and $\gamma$, can be calculated as;
\begin{align}
\alpha = \frac{{6}{{\omega}_s}{A_2}}{{A_{\circ}}^2}
\label{eqn:vdp43}
\end{align}
\begin{align}
\gamma = \frac{{32}{{\omega}_s}{A_3}}{{A_{\circ}}^3}
\label{eqn:vdp44}
\end{align}
External detuning parameter is;
\begin{align}
\sigma = \frac{\Omega - \omega}{\epsilon} \label{eqn:vdp45}
\end{align}
Setting phase $\Gamma$ of the excitation, $\eta$ can be measured
from the phase $\phi$ of ${C_L}(\Omega)$ component in Fourier
Transform of lift signal;
\begin{align}
\eta = \Gamma - {\phi[{C_L}(\Omega)]} \label{eqn:vdp46}
\end{align}
Forcing amplitude ${F_\circ}$ is;
\begin{align}
{F_\circ} = \frac{{2}{{\omega}_s}{A_{\circ}}{\sigma}}{\cos(\eta)}
\label{eqn:vdp47}
\end{align}
Linear damping $\mu$ can be calculated as;
\begin{align}
\mu = {\frac{{\gamma}{{A_{\circ}}^2}}{4}} -
{\frac{{{F_\circ}}{\sin(\eta)}}{{A_{\circ}}{{\omega}_s}}}
\label{eqn:vdp48}
\end{align}
\section{RESULTS \& DISCUSSION}
\label{sec:resultrom} When airfoil starts oscillation in a fluid, a
reverse von Karman vortex street is observed in the wake which leads
to generation of time-varying aerodynamic forces. At low St, these
aerodynamic forces are periodic in nature which may be composed of
several harmonics of fundamental vortex shedding frequency. We
compare the solution of van der Pol oscillator model with those from
CFD, both in temporal and spectral domains. The proposed model not
only captures temporal profiles accurately but also strong
harmonics. in their spectra. We present the validity of the proposed
reduced-order model for three types of forced motions of airfoil;
plunging, pitching and flapping.
\subsection{Plunging Airfoil}
\begin{figure}[!ht]
\centering
\includegraphics[scale=0.6]{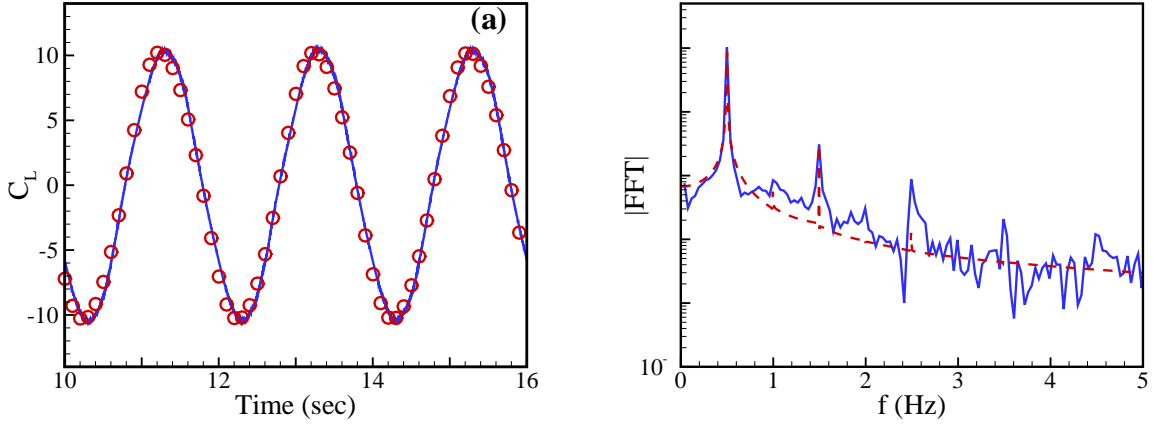}
\caption{Comparison of CFD solution and the proposed model for
Plunging Airfoil at $\mbox{St}=0.45$ (a) Time Histories (Solid
Lines: CFD and Circles: ROM) (b) Amplitude-Spectrum (Solid Lines:
CFD and Dashed Lines: ROM)} \label{fig:CmprsnHeave}
\end{figure}

Plunging airfoil performs oscillatory motion along vertical
direction. It portraits a single-degree of freedom system.
Figure~\ref{fig:CmprsnHeave}a shows comparison of aerodynamic lift
force from CFD simulations and the proposed model in time domain.
Solution for the ordinary differential equation (ODE) of proposed
reduced-order model is obtained by numerical integration using
Runge-Kutta method of order-$4$. It is clear from these figures that
current model fulfills the requisites effectively.
To capture the inherent nonlinear characteristics of lift force, the
results of the proposed model should carry the same spectral
components of the signal. It can be analyzed by comparing the
spectra of signals from CFD and reduced-order model. $C_L$-spectra
of CFD and van der Pol oscillator model are shown in
Fig.~\ref{fig:CmprsnHeave}b. It shows that the proposed model not
only captures the fundamental frequency with accurate amplitude but
also the other harmonic components are well predicted by this
model.
To develop a database for linear and nonlinear damping
parameters, force amplitude, forced and vortex shedding frequencies,
Table~\ref{table2} presents sample dynamic parameters for a range of
St.
\begin{table}[!ht]
\caption{Spectral and ROM Parameters for Lift of Plunging Airfoil}
\begin{center}
\label{table2}
\begin{tabular}{c l l l}
& & \\ 
\hline
Parameters/St & 0.15 & 0.30 & 0.45\\
\hline
$f_s$ & 0.4982 & 0.5003 & 0.4998 \\
$A_\circ$ & 1.908 & 6.051 & 10.3 \\
$A_2$ & 0.001039 & 0.01653 & 0.06951 \\
$A_3$ & 0.08816 & 0.3482 & 0.2933 \\
$\eta$ & $37.0033^\circ$ & $48.2008^\circ$ & $53.0429^\circ$ \\
$\mu$ & 1.1382 & 1.4396 & 0.7144 \\
$\alpha$ & 0.0054 & 0.0085 & 0.0123 \\
$\gamma$ & 1.2714 & 0.1580 & 0.0270 \\
$F_\circ$ & 0.1760 & 0.1513 & 0.0870 \\
\hline
\end{tabular}
\end{center}
\end{table}
\subsection{Pitching Airfoil}
\begin{figure}[!ht]
\centering
\includegraphics[scale=0.6]{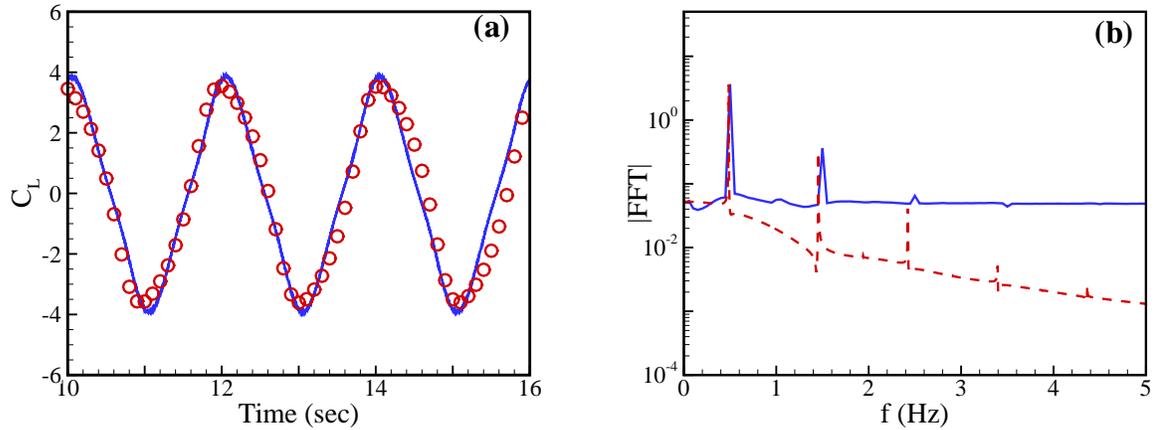}
\caption{Comparison of CFD solution and the proposed model for
Pitching Airfoil at $\mbox{St}=0.30$ (a) Time Histories (Solid
Lines: CFD and Circles: ROM) (b) Amplitude-Spectrum (Solid Lines:
CFD and Dashed Lines: ROM)} \label{fig:CmprsnPitch}
\end{figure}

Although pitching motion of airfoil is a different degree-of-freedom
but its response in terms of aerodynamic forces resembles those of
plunging airfoil. Table~\ref{table3} presents sample values for
spectral and ROM parameters for different values of St. In
Fig.~\ref{fig:CmprsnPitch}, we show comparison of CFD solution with
that of van der Pol oscillator model.
\begin{table}[!ht]
\caption{Spectral and ROM Parameters for Lift of Pitching Airfoil}
\begin{center}
\label{table3}
\begin{tabular}{c l l l}
& & \\ 
\hline
Parameters/St & 0.10 & 0.20 & 0.30\\
\hline
$f_s$ & 0.5002 & 0.5001 & 0.5003 \\
$A_\circ$ & 0.8367 & 2.049 & 3.576 \\
$A_2$ & 0.0005242 & 0.0008243 & 0.002885 \\
$A_3$ & 0.0136 & 0.1203 & 0.3186 \\
$\eta$ & $-0.4177^\circ$ & $6.376^\circ$ & $10.8025^\circ$ \\
$\mu$ & 0.4098 & 1.4757 & 2.2212 \\
$\alpha$ & 0.0141 & 0.0037 & 0.0043 \\
$\gamma$ & 2.3351 & 1.4059 & 0.7008 \\
$F_\circ$ & 0.0072 & 0.0081 & 0.2209 \\
\hline
\end{tabular}
\end{center}
\end{table}
\subsection{Flapping Airfoil}
\begin{figure}[!ht]
\centering
\includegraphics[scale=0.6]{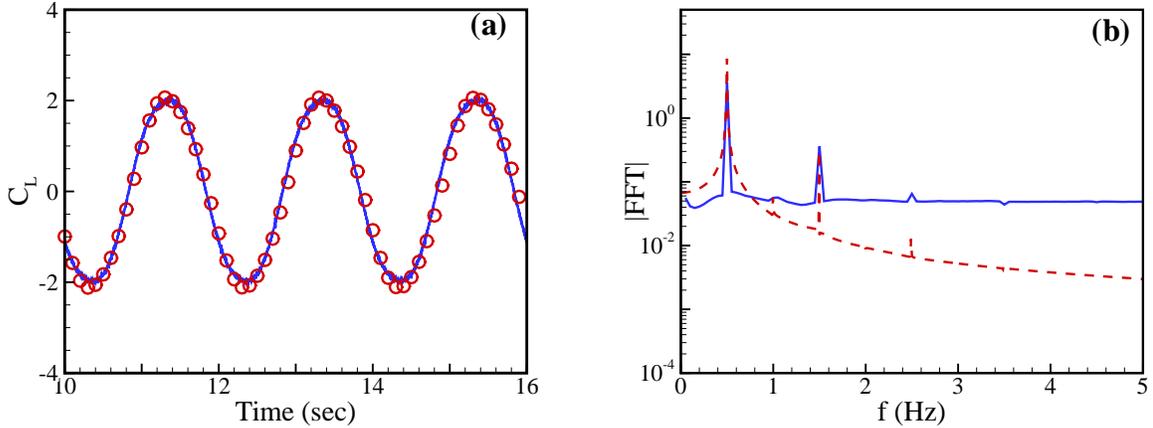}
\caption{Comparison of CFD solution and the proposed model for
Pitching Airfoil at $\mbox{St}=0.20$ (a) Time Histories (Solid
Lines: CFD and Circles: ROM) (b) Amplitude-Spectrum (Solid Lines:
CFD and Dashed Lines: ROM)} \label{fig:Cmprsnflap}
\end{figure}

As described earlier, flapping motion is a combination of plunging
and pitching motions. Practically, swimming and flying species or
robots employ synchronized pitching and plunging while flapping
their wings. Looking at the similar nature of response, we model
lift force of flapping wings using same oscillator model.
Table~\ref{table4} shows sample values of spectral and ROM
parameters for flapping airfoil while Fig.~\ref{fig:Cmprsnflap}
shows comparison of CFD and ROM solutions. CFD simulations presented
here were carried out for $0.05\le h_\circ \le 0.50$,
$\alpha_{\circ}=10^\circ$, $f=0.5\mbox{Hz}$ and $\mbox{Re}=10^3$. It
also proves the suitability of this model for this complex
phenomenon.
\begin{table}[!ht]
\caption{Spectral and ROM Parameters for Lift of Flapping Airfoil}
\begin{center}
\label{table4}
\begin{tabular}{c l l l}
& & \\ 
\hline
Parameters/St & 0.05 & 0.20 & 0.35\\
\hline
$f_s$ & 0.5017 & 0.5003 & 0.4999 \\
$A_\circ$ & 0.6721 & 2.093 & 5.24 \\
$A_2$ & 0.02461 & 0.0288 & 0.04316 \\
$A_3$ & 0.03737 & 0.07439 & 0.1265 \\
$\eta$ & $-31.6684^\circ$ & $60.7861^\circ$ & $60.1451^\circ$ \\
$\mu$ & 1.4077 & 0.8865 & 0.6364 \\
$\alpha$ & 1.0304 & 0.1240 & 0.0456 \\
$\gamma$ & 12.4164 & 0.8161 & 0.0924 \\
$F_\circ$ & 0.0467 & 0.0542 & 0.0450 \\
\hline
\end{tabular}
\end{center}
\end{table}
\section{Improved Model}
\label{sec:imprmod} 
\begin{figure}
\centering
\subfigure[St=0.20]{\label{}\includegraphics[scale=0.5]{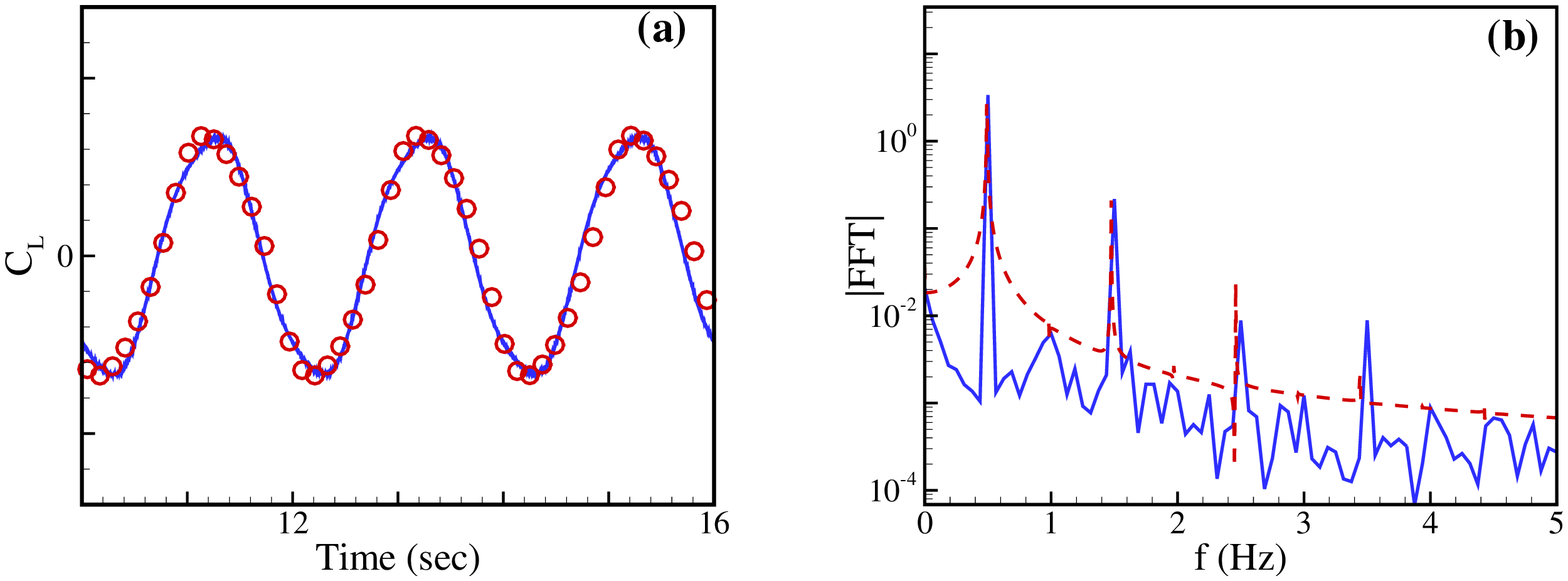}}
\subfigure[St=0.40]{\label{}\includegraphics[scale=0.5]{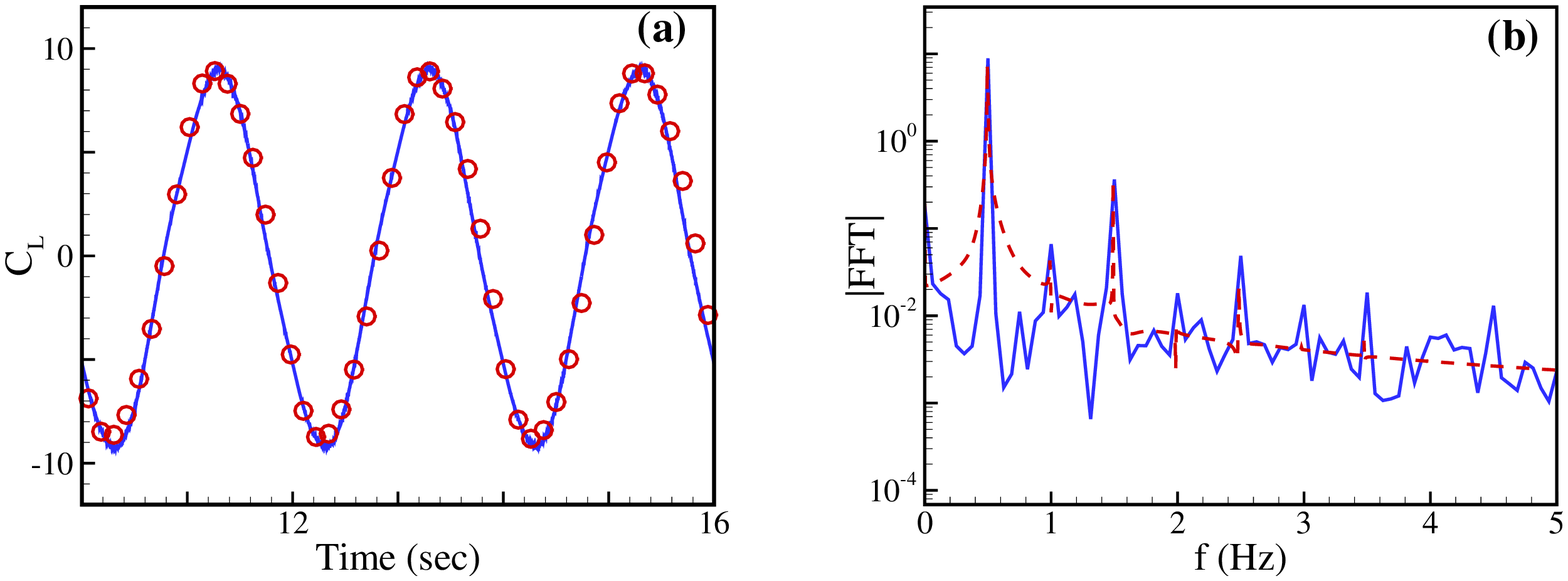}}
\caption{Comparison of solutions from CFD and the improved ROM for
Plunging Airfoil (a) Time Histories (Solid Lines: CFD and Circles:
ROM) (b) Amplitude-Spectrum (Solid Lines: CFD and Dashed Lines:
ROM)} \label{fig:Cmprsnvdp2Heav}
\end{figure}

Like usual unsteady signals, we may decompose
$C_L$ into two components; time-averaged value $\bar{C_L}$ and
fluctuating component $C_l$.
\begin{align}
C_L = \bar{C_L} + C_l
\end{align}
Although motion of airfoil during plunging, pitching and flapping is
symmetrical about its mean position, yet there exists a non-zero
time-averaged value of $C_L$ \cite{Yu2012}. Presence of non-zero
time-averaged value of a signal along with an even harmonic shows
quadratic nonlinearity in the system \cite{Nayfeh1995}. In section
\ref{sec:rom}, we attempt to model the quadratic nonlinearity using
a term ${C_L}{\dot{C_L}}$. It is chosen due to a phase-difference of
$\pi/2$ (or its integral multiple) between fundamental and first
even harmonic in the unsteady $C_L$ signal. This model can capture
the unsteady details effectively but it does not predict $\bar{C_L}$
that is an important feature of aerodynamics in case of oscillating
airfoils. Considering lesser magnitudes of forcing function
amplitude $F_\circ$, a possible singularity at $\eta=\pi/2$ and to
take care of $\bar{C_L}$, we propose another version of the modified
van der Pol oscillator's ordinary differential equation to model the
lift of oscillating airfoil.
\begin{align}
\ddot{C_L}+{{\omega^2}_s}{C_L} &=
{\epsilon}[{\mu}{\dot{C_L}}-{\alpha}{C_L^2}-{\gamma}{{{C_L}^2}{\dot{C_L}}}]
\label{eqn:avdp1}
\end{align}
Solving this system using the method of multiple scales, following
second-order solution is obtained;
\begin{IEEEeqnarray}{rCl}
{C_L}(t) &=& {A_{\circ}}\cos({{\omega}}{t}) + \epsilon[
{\frac{{\alpha}{{A_o}^2}}{2{\omega^2}}} -
{\frac{{\alpha}{{A_{\circ}}^2}}{{6}{{\omega}^2}}}\cos({{2}{\omega{t}+{2{\beta_o}}}}) \nonumber\\
&& -\: {\frac{{\alpha}{{A_{\circ}}^3}}{{{32{\omega}}}}} \cos({{3}
{\omega{t}+{3{\beta_o}}}})] \label{eqn:avdp2}
\end{IEEEeqnarray}
Modulation equations in this case are;
\begin{align}
\dot{A_\circ} = {\mu}{A_o}-{\frac{{\gamma}{{A_o}^3}}{8}}
\label{eqn:avdp3}
\end{align}
\begin{align}
\dot{\beta_\circ} =
-{\frac{\mu^2}{8\omega}}+{\frac{{\mu}{\gamma}{{A_o}^2}}{8\omega}}-{\frac{11{\gamma^2}{{A_o}^4}}{256\omega}}+{\frac{5{\alpha^2}{{A_o}^2}}{12{\omega^3}}}
\label{eqn:avdp4}
\end{align}
Linear and nonlinear damping parameters are identified as;
\begin{align}
\mu = \frac{8\omega{A_3}}{A_o} \label{eqn:avdp6}
\end{align}
\begin{align}
\alpha = \frac{3{\omega^2}{A_2}}{{A_o}^2} \label{eqn:avdp6}
\end{align}
\begin{align}
\gamma = \frac{4\mu}{{A_o}^2} \label{eqn:avdp7}
\end{align}
From Eq.~\ref{eqn:avdp2}, $\bar{C_L}$ is;
\begin{align}
\bar{C_L} = \frac{2\alpha\mu}{\gamma{\omega^2}} \label{eqn:avdp5}
\end{align}
To prove the accuracy of this model, we compare its results with
those from CFD in Fig~\ref{fig:Cmprsnvdp2Heav}. CFD results for
St=0.20 and 0.40 were used here to extract the ROM parameters. The
strength of this model lies in its capability to model time-averaged
value of the lift force. This model, too, behaves well in time as
well as spectral domains. We also show a database for ROM parameters
for different St values in Table~\ref{table5}.
\begin{table}[!ht]
\caption{Parameters for Lift of Plunging Airfoil in Case of Improved
ROM}
\begin{center}
\label{table5}
\begin{tabular}{c l l l}
& & \\ 
\hline
Parameters/St & 0.15 & 0.30 & 0.45\\
\hline
$\mu$ & 1.1613 & 1.4462 & 0.7157 \\
$\alpha$ & 0.0085 & 0.0134 & 0.0194 \\
$\gamma$ & 1.2760 & 0.1580 & 0.0270 \\
$\bar{C_L}$ (ROM) & 0.0046 & 0.0316 & 0.1083 \\
$\bar{C_L}$ (CFD) & 0.0063 & 0.0352 & 0.1190 \\
\hline
\end{tabular}
\end{center}
\end{table}
Comparing damping parameters with those in Table~\ref{table2} for
lift of plunging airfoil, we observe almost equal magnitudes. It is
a manifestation for capability of this improved reduced-order model
to predict actual unsteady characteristics in lift force signal.
\section{Predictive Settings} \label{sec:prdctv} 
\begin{figure}[!ht]
\centering
\includegraphics[scale=0.5]{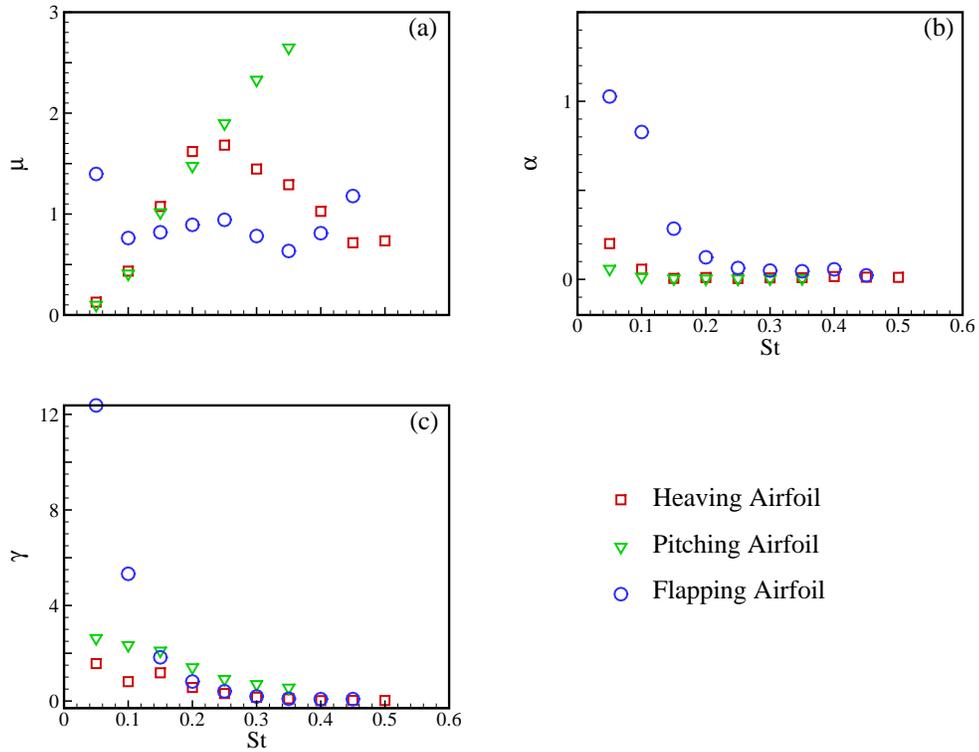}
\caption{Plots of Dynamic Parameters for Varying St}
\label{fig:vdpPrm}
\end{figure}

\begin{figure}[!ht]
\centering
\includegraphics[scale=0.6]{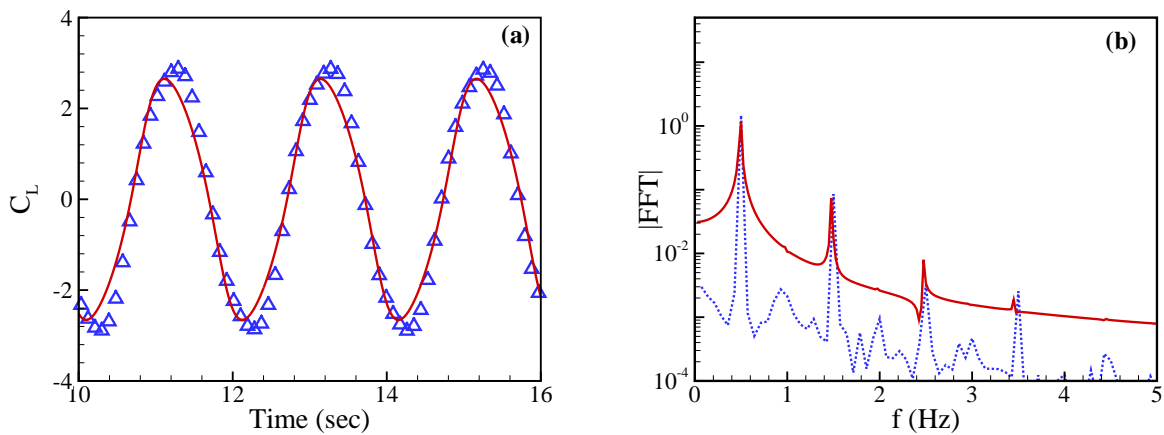}
\caption{Comparison of CFD solution and the proposed model for
Pitching Airfoil at $\mbox{St}=0.20$ (a) Time Histories (Triangles:
CFD and Solid Line: ROM) (b) Amplitude-Spectrum (Dashed Line: CFD
and Solid Lines: ROM)} \label{fig:Prdctv}
\end{figure}

To summarize the
results of the proposed model and its suitability for responses of
various kinematics of airfoil, Fig.~\ref{fig:vdpPrm} shows the
variations of damping coefficients as functions of St. Positive
values of $\mu$, $\alpha$ and $\gamma$ indicate the presence of
limit cycle. Despite a difference in level of their magnitudes in
case of three different kinematics of airfoil, similar trend for
variation can be observed for quadratic ($\alpha$) and cubic
($\gamma$) damping-coefficients. This similar nature of nonlinear
dynamic parameters proves the effectiveness of the present
reduced-order model. Magnitudes of nonlinear damping parameters
decrease due to increasing amplitude $A_\circ$ for fundamental
harmonic as can be seen in Eq.~\ref{eqn:vdp43} and \ref{eqn:vdp44}.
Linear damping $\mu$ for $C_L$ of pitching airfoil comes out to be a
linear function of St while it shows quite similar nonlinear
behavior for plunging and flapping motions.
Now, we examine the performance of our reduced-order models in
predictive settings for $\mbox{St}=0.18$. For this purpose, the
selection of this St value is justified from Fig.~\ref{fig:vdpPrm}
where we observe higher gradients in the trends of both linear and
nonlinear damping parameters. We find out the linear, quadratic and
cubic damping values using data points presented in
Fig.~\ref{fig:vdpPrm} through cubic interpolation scheme. We compute
the solution of the proposed van der Pol oscillator model by
numerical integration and compare its performance with the CFD
results. Figure~\ref{fig:Prdctv} shows that the reduced-order model
well satisfies the requirement not only in time-domain but also in
the spectral domain even at the point around highest variation in
the available data. To quantitatively test the model, we calculate
the percentage errors in the amplitudes of unsteady $C_L$ from CFD
and predicted values of reduced order model, and the fundamental
frequency in the spectra of both. These values come out to be
$7.91\%$ and zero, respectively. While comparing both the spectra,
we see small deviation in the levels of higher harmonics. $C_L$
comes out to be $0.009$ from CFD and the numerical solution of the
model in predictive settings gives $0.006$ which are quite close to
each other. Errors and deviations from actual CFD results may be
removed by using more data points as samples in the interpolation
scheme.
\section{Conclusions}
\label{sec:cnclsion} In the present paper, we identify the existence
of a limit-cycle behavior through aerodynamic forces produced by
oscillating airfoils. It is done through various initial kinematic
states that has greater physical significance. Noting the presence
of even and odd harmonics in $C_L$-spectra for plunging, pitching
and flapping airfoils, we propose a phenomenological reduced-order
model by employing an asymmetric forced van der Pol oscillator
model. Parameters for this models are calculated through results of
CFD simulations. By numerically integrating the governing ROM ODE,
we present comparison of its solutions with those of CFD. Results
are quite promising for both temporal and spectral domains. Real
strength of this proposed model lies in its applicability to
aerodynamic lift forces of plunging, pitching and flapping airfoils
that are altogether different degrees-of-freedom of the relevant
structure. Similarity of nonlinear behavior in all of these
phenomenon are also proven by plotting nonlinear damping
coefficients versus Strouhal number. This model helps measure the
aerodynamic forces of oscillating streamlined body quickly and
accurately without involvement of costly experimental equipment or
time-consuming complex CFD simulations. This model is limited in a
sense that it cannot predict $\bar{C_L}$. To overcome this
deficiency and, considering the lower amplitudes of forcing
functions and a probable singularity condition in forced van der Pol
oscillator model, we propose another model that carries ${C_L}^2$
term to model quadratic nonlinearity with no forcing function. This
model captures $\bar{C_L}$ quite effectively. Magnitudes of linear
and nonlinear dampings in Table~\ref{table4} and ~\ref{table5} are
quite close to each other.
\section*{Acknowledgments}
This work is part of doctoral research of the first author. He is
thankful to National University of Sciences \& Technology and Higher
Education Commission, Government of Pakistan for providing
scholarship under Mega S\&T scheme.
\bibliographystyle{plain}
\bibliography{MyBibFile_Complete}

\begin{thebibliography}{10}

\bibitem{Akhtar2008}
I.~Akhtar.
\newblock {\em {Parallel Simulation, Reduced-Order Modeling and Feedback
  Control of Vortex-Shedding using Fluidic Actuators}}.
\newblock {PhD Thesis}, Virginia Polytechnic Institute and State University,
  USA, 2008.

\bibitem{Akhtar2009A}
I.~Akhtar, O.~A. Marzouk, and A.~H. Nayfeh.
\newblock A van der {P}ol-{D}uffing oscillator model of hydrodynamic forces on
  canonical structures.
\newblock {\em Journal of Computational and Nonlinear Dynamics}, 4(4), 2009.

\bibitem{Ansys}
{ANSYS Fluent}.
\newblock {\em ANSYS Fluent Userguide}.
\newblock ANSYS Inc.

\bibitem{Betz1912}
T.~W. Betz.
\newblock Ein beitrag zur erklarung des segelfluges.
\newblock {\em Zeitschrift fur Flugtechnik und Motorluftschiffahrt},
  3:269--272, 1912.

\bibitem{Fung1998}
J.~Fung.
\newblock {Parameter Identification of Nonlinear Systems Using Perturbation
  Methodsand Higher-Order Statistics}.
\newblock {MS Thesis}, Virginia Polytechnic Institute and State University,
  USA, 1998.

\bibitem{Hajj2000}
M.~R. Hajj, J.~Fung, A.~H. Nayfeh, and S.~O'F. Fahey.
\newblock Damping identification using perturbation techniques and higher order
  spectra.
\newblock {\em Nonlinear Dynamics}, 23:189--203, 2000.

\bibitem{Ho2003}
S.~Ho, H.~Nassef, N.~Pornsinsirirak, Y.~C. Tai, and C.~M. Ho.
\newblock Unsteady aerodynamics and flow control for flapping wing flyers.
\newblock {\em Progress in Aerospace Sciences}, 39:635--681, 2003.

\bibitem{Janajreh2008}
I.~Janajreh and M.~R. Hajj.
\newblock An analytical model for the lift on a rotationally oscillating
  cylinder.
\newblock California, USA, July 2008. BBAA-VI International Colloquium on Bluff
  Bodies Aerodynamics \& Applications.

\bibitem{KhalidJoA2014}
M.~S.~U. Khalid, I.~Akhtar, and N.~I. Durrani.
\newblock Analysis of strouhal number based equivalence of pitching and
  plunging airfoils and wake deflection.
\newblock {\em Proc IMechE Part G: J Aerospace Engineering}, 10.

\bibitem{Knoller1909}
R.~Knoller.
\newblock Gesetze des luftwiderstands.
\newblock {\em Flug-und Motortechnik (Wien)}, 3(21):1--7, 1909.

\bibitem{Lehmann2008}
F.~O. Lehmann.
\newblock Aerial locomotion in flies and robots: Kinematic control and
  aerodynamics of oscillating wings.
\newblock {\em Arthropod Structure and Development}, 33:331--345, 2004.

\bibitem{Marzouk2007}
O.~A. Marzouk, A.~H. Nayfeh, Imran Akhtar, and H.~N. Arafat.
\newblock Modeling steady-state and transient forces on a cylinder.
\newblock {\em Journal of Vibration and Control}, 13(7):1065--1091, 2007.

\bibitem{Mn1998}
F.~C. Moon.
\newblock {\em Applied Dynamics; With Applications to Multibody and Mechatronic
  Systems}.
\newblock John Wiley \& Sons Inc, USA, 1998.

\bibitem{Nayfeh1993}
A.~H. Nayfeh.
\newblock {\em Introduction to Perturbation Techniques}.
\newblock Wiley Classic Library Edition, 1993.

\bibitem{Nayfeh2000}
A.~H. Nayfeh.
\newblock {\em Perturbation Methods}.
\newblock Wiley Classic Library Edition, 2000.

\bibitem{Nayfeh2005}
A.~H. Nayfeh, O.~A. Marzouk, N.~H. Arafat, and I.~Akhtar.
\newblock Modeling the transient and steady-state flow over a stationary
  cylinder.
\newblock In {\em Proceedings of DETC-2005 ASME Design Engineering Technical
  Conference}, 2005.

\bibitem{Nayfeh1995}
A.~H. Nayfeh and D.~T. Mook.
\newblock {\em Nonlinear Oscillations}.
\newblock Wiley Classic Library Edition, 1995.

\bibitem{Nayfeh2003}
A.~H. Nayfeh, F.~Owis, and M.~R. Hajj.
\newblock Model for the coupled lift and drag on a circular cylinder.
\newblock Orlando, Florida, USA, Sep 2003. DETC-2003 ASME 19th Biennial
  Conference on Mechanical Vibration and Noise.

\bibitem{Qin2004}
L.~Qin.
\newblock {\em {Development of Reduced-Order Models for Lift and Drag on
  Oscillating Cylinders with Higher-Order Spectral Moments}}.
\newblock {PhD Thesis}, Virginia Polytechnic Institute and State University,
  USA, 2003.

\bibitem{Shyy2010}
W.~Shyy, H.~Aono, S.~K. Chimakurthi, P.~Trizila, C.~K. Kang, C.~E.~S. Cesnik,
  and H.~Liu.
\newblock Recent progress in flapping wing aerodynamics and aeroelasticity.
\newblock {\em Progress in Aerospace Sciences}, 46:284--327, 2010.

\bibitem{Singh2008}
A.~P. Singh, A.~K. De, V.~K. Carpenter, V.~Eswaran, and K.~Muralidhar.
\newblock Flow past a transversally oscillating square cylinder in free stream
  at low reynolds number.
\newblock {\em International Journal for Numerical Methods in Fluids},
  61(11):658--682, 2008.

\bibitem{Sirovich1987}
L.~Sirovich.
\newblock Turbulence and the dyanmics of coherent structures.
\newblock {\em Quarterly of Applied Mathematics}, 45:561--590, 1987.

\bibitem{Skop1973}
R.~Skop and O.~Griffin.
\newblock A model for the vortex-excited resonant response of bluff cylinders.
\newblock {\em Journal of Sound and Vibration}, 27(2):225--233, 1973.

\bibitem{Skop1975}
R.~Skop and O.~Griffin.
\newblock On a theory for the vortex-excited oscillations of flexible
  cylinderical structures.
\newblock {\em Journal of Sound and Vibration}, 41(3):263--274, 1975.

\bibitem{Skop1997}
R.~A. Skop and S.~Balasubramanian.
\newblock A new twist on an old model for vortex-induced vibrations.
\newblock {\em Journal of Fluids and Structures}, 11:395--412, 1997.

\bibitem{Smith2012}
D.~T. Smith, J.~S. Leontini, J.~Sheridan, and D.~L. Jacono.
\newblock Streamwise forced oscillations of circular and square cylinders.
\newblock {\em Physics of Fluids}, 24(11):11703, 2012.

\bibitem{Triantafyllou2004}
M.~S. Triantafyllou, A.~H. Techet, and F.~S. Hover.
\newblock Review of experimental work in biomimetic foils.
\newblock {\em IEEE Journal of Oceanic Engineering}, 29(3):31--38, July 2004.

\bibitem{Ellenrieder2006}
K.~D. von Ellenrieder.
\newblock Dynamical systems analysis of flapping wing propulsion.
\newblock In {\em Australian Fluid Mechanics Workshop, Melbourne University},
  2006.

\bibitem{Ellenrieder2008}
K.~D. von Ellenrieder, K.~Parker, and J.~Soria.
\newblock Fluid mechanics of flapping wings.
\newblock {\em Experimental Thermal and Fluid Science}, 32:1578--1589, 2008.

\bibitem{Wang2005}
Z.~J. Wang.
\newblock Dissecting insect flight.
\newblock {\em Annual Review of Fluid Mechanics}, 37:183--210, 2005.

\bibitem{Young2007}
J.~Young and J.~C.~S. Lai.
\newblock Vortex lock-in phenoenon in the wake of a plunging airfoil.
\newblock {\em AIAA Journal}, 45(2):485--490, Feb 2007.

\bibitem{Yu2012}
M.~L. Yu, H.~Hu, and Z.~J. Wang.
\newblock Experimental and numerical investigations on the asymmetric wake
  vortex structures around an oscillating airfoil.
\newblock Nashville, Tennessee, USA, Jan 2012. 50th AIAA Aerospace Sciences
  Meeting Including the New Horizons Forum and Aerospace Exposition.
\newblock AIAA-2012-0299.

\end{thebibliography}
%
%
\end{document}